\documentclass[12pt]{article}
\usepackage{amssymb}
\usepackage{amsmath}
\usepackage{graphicx}
\begin{document}
%\markboth{Authors' Names}
%{Slow-roll parameters for inverse quartic potential}
%\catchline{}{}{}{}{}
%
%%%%%%%%%%%%%%%%%%%%%%%%%%%%%%%%%%%%%%%%%%%%%%%%%%%%%%%%%%%%%%%%%%%%

\title{Particle creation and reheating in a braneworld inflationary scenario}

\author{
Neven Bili\'c$^1$\thanks{bilic@irb.hr},  Silvije Domazet$^1$\thanks{sdomazet@irb.hr},
and Goran S.\ Djordjevic$^2$\thanks{gorandj@junis.ni.ac.rs}
   \\
   $^1$Division of Theoretical Physics, Rudjer Bo\v{s}kovi\'{c} Institute, Zagreb, Croatia\\
$^2$Department of Physics,
University of Nis,  Serbia\\
}

\maketitle

\begin{abstract}
 We study  the cosmological particle creation in the tachyon inflation based on the 
 D-brane dynamics in the RSII model extended to include matter in the bulk.
 The presence of matter  modifies the warp factor 
 which results in two effects: 
 a  modification of  the RSII cosmology and
 a  modification of the tachyon potential.
 Besides, a string theory D-brane supports among other fields 
 a U(1) gauge field reflecting open strings attached to the brane. 
 We demonstrate how the interaction of the tachyon with the U(1) gauge field
 drives cosmological creation of massless particles and
 estimate the resulting reheating at the end of inflation. 
 \end{abstract}

%\tableofcontents

\section{Introduction}	

%Among many models of inflation \cite{starobinsky,guth,linde} a popular class comprise tachyon inflation models

The tachyon inflation models \cite{fairbairn,kofman,cline,steer} are a popular class of models
inspired by string theory. 
What distinguishes the tachyon  from the canonical 
scalar field is that the tachyon kinetic term 
is of the Dirac-Born-Infeld (DBI) form \cite{sen}:
\begin{equation}
 \mathcal{L}=-V(\theta) \sqrt{1-g^{\mu\nu}\theta_{,\mu}\theta_{,\nu}} .
\end{equation}
A similar action appears in the so called DBI
inflation models \cite{shandera}. In these models the inflation
is driven by the motion of a D3-brane in a warped throat
region of a compact space and the 
DBI field corresponds to the position of the D3-brane.
One obvious advantage of the tachyon models is that one 
can get around the no-go theorem \cite{ohta} that generally applies to 
string-theory  motivated inflation models. The theorem is
derived under rather  reasonable physical assumption:
absence of higher derivative terms,
non-positivity of the potential, positivity of the canonical kinetic terms for massless fields,
and finiteness  of the
Newton constant.
To have accelerated expansions one has to give up at least one of these
assumptions.
In both tachyon and DBI inflation scenarios
the kinetic term is neither canonical nor positive definite
and hence in this case the no-go theorem does not apply.

Unfortunately, the tachyon inflation suffers from a reheating problem
imminent for all tachyon models with the ground state at 
$\theta\rightarrow\infty$ \cite{kofman}. 
This reheating problem is easily demonstrated for a rather broad class of models
with inverse power law potentials $V(\theta) \propto \theta^{-n}$.
As shown by Abramo and Finelly \cite{abramo}
for $n>2$ 
in the limit $\theta \rightarrow \infty$, $p\rightarrow 0^{-}$ very quickly 
yielding 
a cold dark matter (CDM) domination at the end of inflation.
For $n<2$,  $p\rightarrow -1$ for large $\theta$ and the universe 
behaves as quasi-de Sitter.
After the inflationary epoch in both cases the tachyon will remain a dominant component  
unless at the end of inflation, it
  decayed into inhomogeneous fluctuations and
other particles. This period, known as reheating
\cite{dolgov,brandenberger,kofman2,brandenberger2,kofman3,bassett}, links
the inflationary epoch with the subsequent thermalized
radiation era. In the conventional reheating proposal,
the inflaton field decays perturbatively
into a collection of particles and during the decay it
goes through a large number of oscillations around the
minimum of its potential.  
The tachyon field rolls towards its ground state  
without oscillating about it and the conventional reheating mechanism  does not work.
However, it has been shown \cite{cline} that a coupling of massless fields to the time dependent 
tachyon condensate could yield a reheating efficient enough to overcome the above mention problem
of a CDM dominance. In this paper we explicitly study the reheating that results
from a coupling of the tachyon with a U(1) gauge field.

A simple tachyon model can be analyzed in the framework of the second Randall-Sundrum (RSII)
model \cite{randall2}. The original model consists of two D3-branes in 
the 4+1 dimensional anti de Sitter (AdS$_5$) background 
with line element
\begin{equation}
ds^2_{(5)}=G^{(5)}_{ab} dX^a dX^b=e^{-2|y|/\ell} \eta_{\mu\nu}dx^\mu  dx^\nu 
  -dy^2 ,
 \label{eq3000}
\end{equation}
with the observer brane placed at $y=0$ and the negative tension brane pushed of to $y=\infty$.
 One additional dynamical 3-brane moving in the AdS$_5$ bulk
behaves effectively as a tachyon with a potential
$V(\theta) \propto \theta^{-4}$ and hence it drives a dark matter attractor.
In this paper we  study a braneworld tachyon inflation scenario based on 
a generalized Randall-Sundrum model assuming 
the presence of matter in the bulk, e.g.,  in the form of a minimally coupled scalar field.
This setup is also referred to as a {\em thick brane} \cite{kobayashi,german}.
The bulk scalar will change the braneworld geometry and,
in particular, the braneworld cosmology will differ from that of the original RSII model.
Besides, the tachyon potential, instead of being a simple inverse quartic potential, 
will be a more general function depending on the scalar field self-interaction potential.
In this paper we abbreviate this type of braneworld cosmology by BWC. 

Starting from a given warped geometry 
we can construct the bulk scalar interaction potential and the potential
of the tachyon field that
corresponds to the position of the dynamical D3-brane.
We consider a DBI type effective field theory of rolling tachyon on the D3-brane 
obtained from string theory. 
In particular we analyze a general class of tachyon potentials
and reheating due to the coupling of the tachyon condensate to
the massless abelian gauge field.
We will analyze some typical 
potentials:  a general inverse power law potential and  the exponential potential.

The remainder of the paper is organized as follows. 
In the next section we introduce  the DBI effective field theory and 
derive the density of cosmologically created massless particles. 
In Sec.\ \ref{field} we derive the field equations
in a covariant Hamiltonian formalism.
In Sec.\ \ref{inflation} we discuss the basic equations
of the  tachyon inflation and  estimate the  density of reheating and the density of the tachyons
at the end of inflation.
The concluding remarks are given in  Sec. \ref{conclude}.

\section{Dynamical brane as a tachyon}
\label{dynamical}

The action of the 3+1 dimensional brane in the five dimensional bulk is equivalent 
to the Dirac-Born-Infeld description of a Nambu-Goto 3-brane
\cite{bordemann,jackiw}.
However,
string theory D-branes possess three features that are absent in the simple 
Nambu-Goto membrane action: 
(i) they support an abelian gauge field $A_{\mu}$ reflecting
open strings with their ends stuck on the brane, 
(ii) they couple to the dilaton field $\phi_{\rm d}$,
(iii) they couple to the (pull-back of)
 Kalb-Ramond \cite{kalb20}
antisymmetric tensor field $B_{\mu \nu}$ which, like the gravitational 
field $g_{\mu \nu}$, belongs to the closed string sector.
Consider a $3+1$-dimensional D-brane 
 moving in the 4+1-dimensional bulk spacetime 
  with coordinates $X^{a}$,
$a=0,1,2,3,4$. The points on the brane are parameterized by
$X^a (x^{\mu})$, $\mu=0,1,2,3$, where $x^{\mu}$
are the coordinates on the brane.
In the string frame the action is given by \cite{johnson} 
\begin{equation}
S_{\rm br}= - \sigma
\int d^4x\, e^{-\phi_{\rm d}}\sqrt{-\det (g^{(\rm ind)} + \mathcal{B})}  \, , 
\label{eq0001}
\end{equation}
where $\sigma$ is the brane tension and  $g_{\mu\nu}^{(\rm ind)}$ is the induced metric
or the ``pull back" of the bulk space-time metric 
$G^{(5)}_{ab}$ to the brane,
\begin{equation}
g^{(\rm ind)}_{\mu\nu}=G^{(5)}_{ab}
\frac{\partial X^a}{\partial x^\mu}
\frac{\partial X^b}{\partial x^\nu} \, .
\label{eq0002}
\end{equation}
It will be advantageous to work with line element in conformal coordinates  
\begin{equation}
ds^2_{(5)}=G^{(5)}_{ab} dX^a dX^b=\frac{1}{\chi^2(z)}( g_{\mu\nu}dx^\mu  dx^\nu 
  -dz^2),
 \label{eq4112}
\end{equation}
where the functional form of $\chi(z)$  depends on the selfinteraction potential of the bulk scalar field
\cite{bilic4}.
For a pure AdS bulk 
$\chi=z/\ell$ with $\ell$ being the AdS curvature radius.
We will derive our basic equations assuming an arbitrary monotonously increasing
function of $z$ and specify its form later on when we calculate the reheating.

To derive the induced metric we use
the Gaussian normal parameterization 
 $X^a(x^\mu)=\left(x^\mu, \varTheta\right)$,
with the tachyon field $\varTheta$ substituted for the fifth coordinate $z$
which has become a dynamical field. With this  we find
\begin{equation}
g^{(\rm ind)}_{\mu\nu}=\frac{1}{\chi^2(\varTheta)}
\left( g_{\mu\nu}
  -\varTheta_{,\mu}\varTheta_{,\nu}\right).
 \label{eq2002}
\end{equation}
%where $\eta=\eta(x)$ is the radion field.
The field $\mathcal{B}$ is an antisymmetric 
tensor field that combines
the Kalb-Ramond  and a U(1) gauge fields
$\mathcal{B}_{\mu \nu} = B_{\mu \nu} + 2\pi \alpha' F_{\mu \nu}$.
In the following we will ignore the dilaton and the Kalb-Ramond field $B_{\mu \nu}$.
After a few algebraic manipulations similar to those in Ref.\ \cite{gibbons},
the brane  action may be written as \cite{bilic2} 
 \begin{equation}
S_{\rm br} 
=-\sigma\int d^4 x\sqrt{- g}\:
\chi^{-4}
 \sqrt{(1-X)(1+ Y)
 -Z -W^2 } ,
 \label{eq0006}
\end{equation}
where we have introduced the abbreviations:
%\begin{equation}
%\varphi=(1+ k^2 \varTheta^2\eta)^{-1},
%\end{equation}
\begin{equation}
X=g^{\mu\nu} \varTheta_{,\mu} \varTheta_{,\nu}, 
\quad
W= (2\pi\alpha')^2\chi^4
\frac{\epsilon^{\mu\nu\rho\sigma}F_{\mu\nu}F_{\rho\sigma}}{8\sqrt{-\det g}},
\end{equation}
\begin{equation}
 Y= \frac{(2\pi\alpha')^2}{2}\chi^4F^{\mu\nu}F_{\mu\nu},
\quad
Z=\frac{(2\pi\alpha')^2}{2} \chi^4
\varTheta_{,\mu}{F^\mu}_\nu 
F^{\nu\rho}\varTheta_{,\rho} \, .
\nonumber
\end{equation}
Next, neglecting the dilaton, expanding (\ref{eq0006}), and keeping the terms up to quadratic  order
in $F_{\mu\nu}$ we obtain
\begin{equation}
S_{\rm br} 
=-\sigma\int d^4 x\sqrt{-g}\:
\left(
\frac{\sqrt{1-X}}{\chi^{4}} +\frac{(2\pi\alpha')^2}{4}F^{\mu\nu}F_{\mu\nu}
  \right)+S_{\rm int} ,
 \label{eq0007}
\end{equation}
where
\begin{equation}
S_{\rm int} 
=\sigma \frac{(2\pi\alpha')^2}{2} \int d^4 x\sqrt{-g}\:
 \left( F^{\mu}_{\rho}F^{\rho\nu}\varTheta_{,\mu}\varTheta_{,\nu}
 +\frac{X}{4}F^{\mu\nu}F_{\mu\nu}\right) . 
 \label{eq0017}
\end{equation}
The first term in brackets in (\ref{eq0007}) is the basic 
tachyon Lagrangian with potential 
\begin{equation}
V(\varTheta)=\sigma\chi^{-4}(\varTheta) , 
\end{equation}
the second term is the Maxwell Lagrangian 
provided 
$\sigma=(2\pi \alpha')^{-2}$. 
The action (\ref{eq0017}) describes  the  interaction between
the tachyon and the gauge field 
which will be responsible for reheating at the end of inflation.
It is convenient to express this term in the form 
 \begin{equation}
S_{\rm int} 
=\sigma(2\pi\alpha')^2 \int d^4 x\left[\sqrt{-g}
 \frac{f(X)}{2}
 \frac{F^{\mu\nu}F_{\mu\nu}}{4}
 -\sqrt{- G}
 G^{\mu\alpha}G^{\nu\beta}
 \frac{F_{\mu\nu}F_{\alpha\beta}}{4}
 \right].
  \label{eq0004}
\end{equation}
where
\begin{equation}
f(X)= 
 \sqrt{4+ X^2}
   \label{eq0005} .
\end{equation}
In (\ref{eq0004}) we have introduced the effective metric tensor and its determinant 
respectively as \cite{tolic1,tolic2}
\begin{equation}
G_{\mu\nu}=\Omega_{\rm c}
\left[g_{\mu\nu}-(1-c_{\rm s}^2)u_\mu u_\nu \right]\, ,
\label{eq3008}
\end{equation}
\begin{equation}
 G\equiv \det G_{\mu\nu}= \Omega_{\rm c}^4 c_{\rm s}^2 g ,
\end{equation}
 where
\begin{equation}
 u_\mu=\frac{\varTheta_{,\mu}}{\sqrt{X}},  \quad 
  u^\mu=\frac{g^{\mu\nu}\varTheta_{,\nu}}{\sqrt{X}}
\end{equation}
are the components of the tachyon fluid four-velocity,
$\Omega_{\rm c}$ is an arbitrary conformal factor, and $c_{\rm s}$ is
the effective sound speed defined by
\begin{equation}
 c_{\rm s}^2=\frac{f(X)-X}{f(X)+X}.
\end{equation}
The last term in square brackets in (\ref{eq0004}) being conformally invariant 
will not contribute to the creation of photons \cite{parker4}.
Hence, the reheating will be affected only by the first term in square brackets
due to the time dependent factor  in front of the  Maxwell Lagrangian.

To simplify the estimate of the photon creation rate we will replace the Maxwell term
$F^{\mu\nu}F_{\mu\nu}$ by the Lagrangian for
two noninteracting massless scalar degrees of freedom conformally coupled to gravity.
With this simplification  
the interaction term is represented by a free scalar field in an effective
time dependent gravitational field and we can estimate the creation rate using the
well known method of adiabatic expansion \cite{parker1}.
For each scalar degree of freedom $\varphi$ we  take  
 \begin{equation}
S_{\rm scal} 
=\frac{\sigma(2\pi\alpha')^2}{2} \int d^4 x\sqrt{-g}
 \frac{f(X)}{2}
 \left(g^{\mu\nu}\varphi_{,\mu}\varphi_{,\nu}-\frac16 R \varphi^2\right),
   \label{eq0009}
\end{equation}
where $R$ is the Ricci scalar associated with the metric $g_{\mu\nu}$.

In the following we will assume the background metric to be spatially flat FRW spacetime 
with four dimensional line element in the form 
\begin{equation}
 ds^2=g_{\mu\nu}dx^\mu dx^\nu=dt^2-a^2(t)(dr^2+r^2 d\Omega^2) .% \mbox{\boldmath$dx$}^2
 \label{eq0012}
\end{equation}
%Note that, unlike in Sec. \ref{braneworld}, the time $t$ here is synchronous.
In the cosmological context it is natural to assume that the
tachyon condensate is comoving, i.e., the velocity components are $u_\mu=(1,0,0,0)$.
 Then $f$ becomes a function of $\dot{\varTheta}$ only
\begin{equation}
 f=\sqrt{4+\dot{\varTheta}^4} .
 \label{eq2036}
\end{equation}
By way of a conformal transformation 
\begin{equation}
\tilde{g}_{\mu\nu}=fg_{\mu\nu}, \quad 
\tilde{g}^{\mu\nu}=f^{-1}g^{\mu\nu}, \quad
\tilde g=\det \tilde{g}_{\mu\nu}= f^4 g ,
   \label{eq0011}
\end{equation}
the action may be expressed in the form
\begin{equation}
S_{\rm scal} 
=\int d^4 x\sqrt{-\tilde{g}}
 \frac12 \left[ \tilde{g}^{\mu\nu}\varphi_{,\mu}\varphi_{,\nu}
 -\left(\frac{\tilde{R}}{6}+m_{\rm eff}^2\right)\varphi^2  \right] ,
   \label{eq0020}
\end{equation}
where 
  \begin{equation}
 \tilde{R}=\frac{R}{f} +\frac{6}{f}\left[
 \frac32 \frac{\dot{a}\dot{f}}{af}+\frac12 \frac{\ddot{f}}{f}-\frac14\left(\frac{\dot{f}}{f}\right)^2
 \right].
 \label{eq2110}
 \end{equation} 
 is the Ricci scalar associated with the metric $\tilde{g}_{\mu\nu}$,
 and 
  \begin{equation}
 m_{\rm eff}^2=\frac{1}{f}\left[\frac14\left(\frac{\dot{f}}{f}\right)^2
 -\frac32 \frac{\dot{a}\dot{f}}{af}-\frac12 \frac{\ddot{f}}{f}
 \right].
 \label{eq2111}
 \end{equation} 
is a time dependent effective mass squared.
In (\ref{eq0020}) we have omitted the constant factor $\sigma(2\pi\alpha')^2/2$
as it does not affect the cosmological particle creation.

The energy density of created particles for two massless  fields is  given by
\begin{equation}
 \rho_{\rm rad}(t) =
\frac{1}{\pi^2a^3}\int_0^\infty  dq q^2\omega(t)|\beta_q(t)|^2=
 \frac{1}{\pi^2}\int_0^\infty d\omega \omega^3|\beta_q(t)|^2,
 \label{eq2031}
\end{equation}
where $\omega=q/a(t)$ is the time dependent frequency equal to the physical momentum.
The quantity $|\beta_q(t)|^2$ is the square of the Bogoliubov coefficient 
which represents the spectral density of particles created at time $t$.
For a fixed  comoving momentum $q$  
the square of the Bogoliubov coefficient at an arbitrary time $t$ is given by \cite{tolic2} 
\begin{equation}
 |\beta_q|^2= \frac14 \left[\frac{\omega}{W}+\frac{W}{\omega}
 +\frac{1}{4\omega W}\left( \frac{\dot{W}}{W}-\frac{\dot{\omega}}{\omega}\right)^2 -2\right],
 \label{eq2033}
\end{equation}
where
%the prime $'$ denotes a derivative with respect to $t$ and
the positive function $W(t)$ satisfies the differential equation
\begin{equation}
W^2=\Omega^2+ W^{1/2}\frac{d^2}{dt^2}(W^{-1/2}), 
\label{eq2014}
\end{equation}
with initial conditions 
%\begin{equation}
$W(t_0)=\omega(t_0)$ and $\dot{W}(t_0)=\dot{\omega}(t_0)$
% \label{eq2034}
%\end{equation}
at a conveniently chosen $t_0$, e.g., at the beginning of inflation.
The time-dependent function $\Omega$ is given by
\begin{equation}
\Omega^2=\omega^2+ f m_{\rm eff}^2 
+\frac14
\left(\frac{\dot{a}}{a}\right)^2-\frac12
\frac{\ddot{a}}{a} .
\label{eq2010}
\end{equation}

 {\em Nota bene} :
 If  the effective mass $m_{\rm eff}$ were equal to zero 
 then, as  may be easily verified,
 the function $W=\omega$ would be a solution to (\ref{eq2014})
 for an arbitrary $a$ and, by virtue of (\ref{eq2033}), $|\beta_q|^2$ would vanish
 identically.  
  This is to be expected since in this case 
 the action (\ref{eq2110})
would describe  a conformally coupled massless scalar field and hence there would be no
particle creation.

As usual, we expect the integral (\ref{eq2031}) to diverge at the upper bound.
To check the UV limit of the integrand 
we need the asymptotic expression for $ |\beta_q|^2$.
The behavior of $ |\beta_q|^2$  in the limit $q \rightarrow \infty$ is obtained from Eq.\ (\ref{eq2033}) 
by making use of the second order adiabatic expansion  
\begin{equation}
 W=\omega+ \omega^{(2)}+ \mathcal{O}(\omega^{-3}).
 \label{eq2017}
\end{equation}
Applying the general result of Ref.\ \cite{tolic2}
to our expression (\ref{eq2010}) we find
\begin{equation}
 \omega^{(2)}={2\omega} f m_{\rm eff}^2.
 \label{eq2012}
\end{equation}
Then from (\ref{eq2033}) and (\ref{eq2017}) with (\ref{eq2012}) we obtain
\begin{equation}
|\beta_q|^2=
 \frac{1}{16} \frac{F^4 }{\omega^4} + \mathcal{O}(\omega^{-6}) ,
\label{eq2015}
\end{equation}
where
\begin{equation}
F^2=\left|
 \frac32 \frac{\dot{a}\dot{f}}{af}+\frac12 \frac{\ddot{f}}{f}-\frac14\left(\frac{\dot{f}}{f}\right)^2
 \right|.
\label{eq2016}
\end{equation}
Hence,  the integral in (\ref{eq2031}) diverges logarithmically.
However, we know that the estimate of the spectral functions is unreliable
beyond the string scale so 
 we may choose the cutoff of the order of $M_{\rm s}=1/\sqrt{\alpha'}$.
 In practice one can do the integral up to some large momentum using the exact function (\ref{eq2033}) 
 and the remainder of the integral estimate using the asymptotic expression 
 (\ref{eq2015}).

 To evaluate the energy density (\ref{eq2031}) and compare it with the tachyon energy density at the end of inflation
 we need to study the evolution of the tachyon fluid during inflation. This  will be done in the next section.

\section{Field equations}
\label{field}
In this section we derive the tachyon field equations 
from  the  action (\ref{eq0007}) ignoring the gauge field.
Then, the tachyon Lagrangian takes the form 
\begin{equation}
{\cal{L}} =  
-\frac{\sigma}{\chi(\varTheta)^4}\sqrt{1-g^{\mu\nu}\varTheta_{,\mu}\varTheta_{,\nu}}  \, .
 \label{eq000}
\end{equation}

In the following we will assume the spatially flat FRW spacetime on the observer brane
with four dimensional line element in the standard form (\ref{eq0012}).
The treatment of our system in a cosmological context is conveniently performed 
in the covariant Hamiltonian formalism
\cite{dedonder,struckmeier,bilic}.
To this end we first define 
the conjugate momentum field as
\begin{equation}
\Pi_\varTheta^\mu=
\frac{\partial{\cal{L}}}{\partial\varTheta_{,\mu}}  .
\end{equation}
In the cosmological context  $\Pi_\varTheta^\mu$ is time-like so we may also define 
its magnitude as 
\begin{equation}
\Pi_\varTheta=\sqrt{g_{\mu\nu}\Pi_\varTheta^\mu\Pi_\varTheta^\nu} \, .
\label{eq2118}
\end{equation}
The Hamiltonian density may  be derived from the stress tensor corresponding to the
Lagrangian (\ref{eq000}) or by the Legendre transformation.
Either way one finds \cite{bilic}
\begin{equation}
{\cal{H}} =\frac{\sigma}{\chi^4}\sqrt{1+\Pi_{\varTheta}^2\chi^8/\sigma^2}.
 \label{eq001}
\end{equation}
Then, we can write Hamilton's equations in the form
\begin{eqnarray}
%\dot{\varPhi} = \frac{\partial{\cal{H}}}{\partial\Pi_\varPhi},\label{eqHam1}\\
\dot{\varTheta} = \frac{\partial{\cal{H}}}{\partial\Pi_\varTheta},\label{eqHam2} \\
%\dot{\Pi}_\varPhi +3H\Pi_\varPhi=-\frac{\partial{\cal{H}}}{\partial\varPhi},\label{eqHam3}\\
\dot{\Pi}_\varTheta + 3H\Pi_\varTheta=-\frac{\partial{\cal{H}}}{\partial\varTheta}.
\label{eqHam4}
\end{eqnarray}
In the spatially flat BWC
the Hubble expansion rate $H$ is related to the Hamiltonian via 
a modified Friedmann equation \cite{bilic4} which can be written as
\begin{equation} 
 H\equiv\frac{\dot{a}}{a}=\sqrt{\frac{8 \pi G_{\rm N}}{3} \mathcal{H}\left(\frac{\chi_{,\varTheta}}{k}
+ \frac{2 \pi G_{\rm N}}{3k^2} \mathcal{H}\right) }.
% \nonumber
\label{scale_a}
\end{equation}
where $k=G_{\rm N}/G_5$ is a mass scale which will later be fixed from phenomenology, and 
%\begin{equation}
$\chi_{,\varTheta}$  
% \label{eq3100}
%\end{equation}
 is an abbreviation for $\partial\chi/\partial\varTheta$.
 In addition, we will make use of the energy-momentum  conservation equation
combined with the time derivative of (\ref{scale_a}) to obtain
the second Friedmann equation
\begin{equation}
\dot{H}=-4\pi G_{\rm N}(\mathcal{H}+\mathcal{L})\left(\frac{\chi_{,\varTheta}}{k}+
\frac{4 \pi G_{\rm N}}{3k^2} \mathcal{H}\right)
+\sqrt{\frac{2 \pi G_{\rm N}\mathcal{H}}{3k \chi_{,\varTheta}+2\pi G_{\rm N}\mathcal{H} }}\dot{\chi}_{,\varTheta}\, .
 \label{eq3222}
\end{equation}
In the pure AdS bulk $\chi_{,\varTheta}=1$ in which case 
 one recovers the usual RSII modifications of the Friedmann equations \cite{binetruy}.
 %Thus, the Friedman equations are modified in the RSII cosmology.
 
The last term on the righthand side of  Eq.\ (\ref{eq3222}) containing the time derivative $\dot{\chi}_{,\varTheta}$
could be neglected provided
\begin{equation}
\frac{|\dot{\chi}_{,\varTheta}|}{\chi_{,\varTheta}} \ll \frac{\dot{|\mathcal{H}}|}{\mathcal{H}}.
\label{eq3223}
\end{equation}
It may be easily shown that this approximation  
 is justified as long as 
\begin{equation}
 \frac{\chi\left|\chi_{,\varTheta\varTheta}\right|}{\chi_{,\varTheta}^2} \ll 4 ,
 \label{eq3224}
\end{equation}
which may be checked once the function $\chi(\varTheta)$ is specified.
For example, in 
the original RSII model where
$\chi(\varTheta) =k\varTheta$ the inequality (\ref{eq3224}) is trivially satisfied.
For an exponential dependence
$\chi(\varTheta)\propto e^{k\varTheta}$  and 
a general power law $\chi\propto \varTheta^n$, $n\gtrsim 1/4$, one finds
$\chi|\chi_{,\varTheta\varTheta}|/\chi_{,\varTheta}^2=1$ and 
$|1-1/n|$, respectively, so 
in these two cases 
Eq.\ (\ref{eq3224}) is
marginally satisfied.

To solve the system of equations (\ref{eqHam2})-(\ref{scale_a}) 
it is convenient to rescale the time as $t=\tau/k$ and  express  
the system in terms of dimensionless quantities. 
To this end  we  introduce  the dimensionless functions 
\begin{eqnarray}
h = H/k, 
\quad
\theta=k \varTheta, \quad
\pi_\theta = \Pi_{\varTheta}/\sigma.
\label{eq002}
\end{eqnarray}
%and using 
Besides, we rescale the Lagrangian and Hamiltonian to obtain the 
rescaled dimensionless
pressure and energy density:
\begin{equation}
 p= \frac{\mathcal{L}}{\sigma}=-\frac{1}{\chi^4\sqrt{1+\chi^8\pi_{\theta}^2}}=
 -\frac{1}{\chi^4}\sqrt{1-\dot{\theta}^2} ,
 \label{eq0081}
\end{equation}
\begin{equation}
 \rho= \frac{\mathcal{H}}{\sigma}=\frac{1}{\chi^4}\sqrt{1+\chi^8\pi_{\theta}^2}=
 \frac{1}{\chi^4}\frac{1}{\sqrt{1-\dot{\theta}^2}}.
 \label{eq008}
\end{equation}
In these equations  and from now on the overdot denotes a derivative with respect to $\tau$.
Then, following Ref.\ \cite{bilic3}, we introduce a dimensionless coupling
\begin{equation}
\kappa^2=\frac{8\pi G_{\rm N}}{k^2}\sigma =\frac{8\pi G_5}{G_{\rm N}}\sigma
\label{eq102}
\end{equation}
and from (\ref{eqHam2})-(\ref{scale_a})
we obtain the following set of equations 
\begin{equation}
\dot \theta=\frac{\chi^4\pi_{\theta}}
{\sqrt{1+\chi^8\pi_{\theta}^2}}
=\frac{\pi_{\theta}}{\rho} ,
\label{eq003}
\end{equation}
\begin{equation}
\dot \pi_\theta=-3h\pi_\theta
+\frac{4\chi_{,\theta}}{\chi^5
\sqrt{1+\chi^8\pi_\theta^2}},
\label{eq004}
\end{equation}
where
\begin{eqnarray}
\label{h}
h=\sqrt{\frac{\kappa^2}{3}\rho\left(\chi_{,\theta}+\frac{\kappa^2}{12}\rho  \right)}.
\end{eqnarray}
In addition,  from (\ref{eq3222}) we obtain  
the second Friedman equation in dimensionless form
\begin{eqnarray}
\dot{h}=-\frac{\kappa^2}{2}(\rho+p)\left(\chi_{,\theta}+\frac{\kappa^2}{6}\rho  \right)
+\frac{\kappa^2\rho}{6h}\chi_{,\theta\theta}\dot{\theta} .
\label{h2}
\end{eqnarray}
Obviously, the explicit dependence on $\sigma$ and $k$ in Eqs.\ (\ref{eq003})-(\ref{h2}) is eliminated 
leaving one dimensionless
 free parameter $\kappa$.
%
%

%
%together with the Friedmann equation for the scale $a(t)$

\section{Tachyon inflation and reheating}
\label{inflation}

%\subsection{Slow-roll parameters}
The basic quantities in all inflation models are
the so called slow-roll parameters defined as \cite{steer,schwarz} 
\begin{equation}
\epsilon_{i} \equiv \frac{d\ln| \epsilon_{i-1}|}{Hdt},\qquad i \geq 1,
\label{eq4100}
\end{equation}
where
\begin{equation}
\epsilon_0 \equiv \frac{H_*}{H} 
\end{equation}
and $H_*$ is the Hubble rate at an arbitrarily chosen time. 
The conditions for a slow-roll regime are satisfied when $\epsilon_1 < 1$ and $\epsilon_2 < 1$, 
and inflation ends when any of them exceeds unity.
%The effect of the radion and the tachyon can be seen if we compare 
%the slow-roll parameters for the full model with those for the model with inverse quartic tachyon potential.

%
%\subsection{Slow-roll approximation}
%

Tachyon inflation is based upon the slow evolution of $\theta$ with the slow-roll
conditions \cite{steer}
\begin{equation}
 \dot{\theta}^2\ll 1, \quad \ddot{\theta} \ll 3h \dot{\theta}.
 \label{eq4101}
\end{equation}
It may be shown that, in our formalism, the slow-roll conditions
equivalent to   (\ref{eq4101}) are
\begin{equation}
 \dot{\theta}\simeq \chi^4\pi_\theta \ll 1, \quad \dot\pi_\theta \ll 3h \pi_\theta ,
 \label{eq1001}
\end{equation}
so that in the slow-roll approximation we may  neglect the factors $(1-\dot{\theta}^2)^{-1/2}=
(1+\chi^8\pi_\theta^2)^{1/2}$ in
(\ref{eq0081}) and (\ref{eq008}).
Then, during inflation we have
\begin{equation}
h\simeq \frac{\kappa}{\sqrt3 \chi^2}\left(\chi_{,\theta}+\frac{\kappa^2}{12\chi^4}\right)^{1/2},
\label{eq1007}
\end{equation}
\begin{equation}
 \dot{\theta} \simeq \frac{4\chi_{,\theta}}{3h\chi} \simeq \frac{4\chi \chi_{,\theta}}{\sqrt3 \kappa}
\left(\chi_{,\theta}+\frac{\kappa^2}{12\chi^4}\right)^{-1/2} ,
\label{eq1008}
\end{equation}
and
\begin{equation}
 \ddot{\theta} \simeq \frac{4\dot{\theta}}{\sqrt3 \kappa}
 \left(\chi_{,\theta}+\frac{\kappa^2}{12\chi^4}\right)^{-3/2}
 \left[\chi_{,\theta}^2 
 \left(\chi_{,\theta} +\frac{\kappa^2}{4\chi^4}\right)
+ \frac12 \chi\chi_{,\theta\theta}\left(\chi_{,\theta} +\frac{\kappa^2}{6\chi^4}\right)
\right].
\label{eq1009}
\end{equation}
As a consequence, the first two slow-roll parameters defined in (\ref{eq4100}) can be approximated by  
\begin{eqnarray}
\epsilon_1 = -\frac{\dot{h}}{h^2}\simeq
%\frac32 \dot{\theta}^2\left(\chi_{,\theta}+\frac{\kappa^2}{6\chi^4}\right)
%\left(\chi_{,\theta}+\frac{\kappa^2}{12\chi^4}\right)^{-1}
%\nonumber \\ 
%\simeq
\frac{8\chi^2\chi_{,\theta}^2}{\kappa^2}\left(\chi_{,\theta}+
\frac{\kappa^2}{6\chi^4}-\frac{\chi\chi_{,\theta\theta}}{4\chi_{,\theta}}\right)
\left(\chi_{,\theta}+\frac{\kappa^2}{12\chi^4}\right)^{-2} ,
\label{eq009}
\end{eqnarray}
\begin{eqnarray}
\epsilon_2 =2\epsilon_1+ \frac{\ddot{h}}{h\dot{h}}
%&\simeq& 2\frac{\ddot{\theta}}{h\dot{\theta}}
%-\dot{\theta}^2\frac{\kappa^2\chi_{,\theta}}{4\chi^4}
%\left(\chi_{,\theta}+\frac{\kappa^2}{6\chi^4}\right)^{-1}\left(\chi_{,\theta}+\frac{\kappa^2}{12\chi^4}\right)^{-1}
%\nonumber \\ 
\simeq 
 \frac{8\chi^2\chi_{,\theta}^2}{\kappa^2} \left(\chi_{,\theta}+\frac{\kappa^2}{12\chi^4}\right)^{-2}\left[
\chi_{,\theta}+\frac{\kappa^2}{4\chi^4}-\frac{\kappa^2\chi_{,\theta}}{6\chi^4}
\left(\chi_{,\theta}+\frac{\kappa^2}{6\chi^4}\right)^{-1}
\right].
\label{eq010}
\end{eqnarray}
This  should be contrasted with the corresponding results of the  tachyon inflation
in the standard cosmology:
\begin{equation}
 \epsilon_1 \simeq \epsilon_2 \simeq\frac{8\chi^2\chi_{,\theta}^2}{\kappa^2} .
 \label{eq011}
\end{equation}
In Eq. (\ref{eq011}) and in the expression for $\epsilon_2$ in (\ref{eq010})
we have neglected the contribution of the 
terms proportional to the second derivative $\chi_{,\theta\theta}$.

Close to and at the end of inflation $\chi(\theta)\gg 1$  so  the contribution of the inverse quartic term 
$\kappa^2/\chi^4$ will be negligible compared to $\chi_{,\theta}$. Then 
\begin{equation}
h\simeq \frac{1}{\sqrt{3}} \frac{\kappa\chi_{,\theta}^{1/2}}{\chi^2},
\quad\quad
\dot{\theta}\simeq \frac{4}{\sqrt{3}} \frac{\chi\chi_{,\theta}^{1/2}}{ \kappa},
\label{eq0021}
\end{equation}
\begin{equation}
\ddot{\theta}\simeq \frac{4}{\sqrt{3}\kappa}
\left( \chi_{,\theta}^{3/2}+ \frac12 \chi \chi_{,\theta}^{-1/2}\chi_{,\theta\theta}\right)\dot{\theta},
 \label{eq0022}
\end{equation}
\begin{equation}
\epsilon_1\simeq 
%\frac32 \dot{\theta}^2\simeq 
8 \frac{\chi^2\chi_{,\theta}}{\kappa^2}\left(1- \frac{\chi\chi_{,\theta\theta}}{4 \chi_{,\theta}^2}
\right),
 \label{eq0018}
\end{equation}
and $\epsilon_2\simeq \epsilon_1$ also holds if we neglect the contribution of 
the $\chi_{,\theta\theta}$ term in (\ref{eq0018}). Hence, in the slow-roll regime the tachyon inflation in the BW modified cosmology proceeds in a quite distinct
way compared with that in the standard FRW cosmology. 
The expressions (\ref{eq0021}) and (\ref{eq0018})
can be used at the end of inflation where one requires 
\begin{equation}
 \epsilon_1(\theta_{\rm f})\simeq 
 8\frac{\chi^2(\theta_{\rm f})\chi_{,\theta}(\theta_{\rm f})}{\kappa^2}
 \left(1- \frac{\chi(\theta_{\rm f})\chi_{,\theta\theta}(\theta_{\rm f})}{4 \chi_{,\theta}^2(\theta_{\rm f})}
\right)
 \simeq 1 .
\label{eq017} 
 \end{equation}
 
Unlike the end of inflation, the beginning of inflation is characterized by 
$\chi_{,\theta}(\theta_i) \ll\kappa^2/(12\chi^4(\theta_{\rm i}))$,
hence, the $\chi_{,\theta}(\theta_i)$ may be neglected with respect to $\kappa^2/(12\chi^4(\theta_{\rm i}))$.
Besides, the terms proportional to  $\chi_{,\theta\theta}$ 
may also be neglected so we find
\begin{equation}
\epsilon_1(\theta_{\rm i})\simeq 192 \frac{\chi^6(\theta_{\rm i})\chi_{,\theta}^2(\theta_{\rm i})}{\kappa^4}
\quad\quad
\epsilon_2(\theta_{\rm i})\simeq 288 \frac{\chi^6(\theta_{\rm i})\chi_{,\theta}^2(\theta_{\rm i})}{\kappa^4}
\simeq \frac32\epsilon_1(\theta_{\rm i}) .
 \label{eq018}
\end{equation}
Then, in the slow-roll approximation  the number of $e$-folds  is given by 
\begin{equation}\label{inteq}
N \simeq\frac{\kappa^2}{4} \int_{\chi_{\rm i}}^{\chi_{\rm{f}}} \frac{d\chi}{\chi^3\chi_{,\theta}^2}
\left(\chi_{,\theta}+\frac{\kappa^2}{12\chi^4}\right) .
%\simeq
%\frac{\kappa^2}{8\theta_{\rm i}^2}\left(\chi_{,\theta}+\frac{\kappa^2}{36\theta_{\rm i}^4}\right)-1
%\simeq\frac{\kappa^4}{288\theta_{\rm i}^6}-1,
%\simeq \frac{1}{\epsilon_1(\theta_{\rm i})}
%\left(1+\frac{\kappa^2}{6\theta^4}\right)\left(1+\frac{\kappa^2}{36\theta^4}\right)
%\left(1+\frac{\kappa^2}{12\theta^4}\right)^{-2}.
\end{equation}
The subscripts  ${\rm i}$ and ${\rm f}$  
in Eqs. (\ref{eq017})-(\ref{inteq}) denote the beginning and the end of inflation, respectively.
Specifically, 
for the exponential potential, i.e., for  $\chi(\theta)=e^{\theta/4}$ 
\begin{equation}
N_{\rm exp} \simeq
\frac{\kappa^4}{24\chi_{\rm i}^8}\left(1+8\frac{\chi_{\rm i}^5}{\kappa^2}\right)-\frac23
\simeq\frac{1}{2\epsilon_1(\theta_{\rm i})}-\frac23,
\label{eq0032}
\end{equation}
whereas for a general power law potential, i.e., for  $\chi(\theta)=\theta^n$, with $n>1/4$ and $n\neq1/3$,
we obtain
\begin{eqnarray}
&&N_n \simeq
\frac{\kappa^4}{96 n (4n-1)\chi_{\rm i}^{8-2/n}}\left(1+\frac{24n(4n-1)}{3n-1}
\frac{\chi_{\rm i}^{5-1/n}}{\kappa^2}\right)-\frac{3n+1}{2(3n-1)} 
\nonumber \\
&&
\simeq\frac{2n}{(4n-1)\epsilon_1(\theta_{\rm i})}-\frac{3n+1}{2(3n-1)} .
\label{eq0033}
\end{eqnarray}
 
%\subsection{Order of magnitude estimate} 
The tachyon energy density is obtained by multiplying 
(\ref{eq008}) by $\sigma$, i.e.,
\begin{equation}
\rho_{\rm tach}= \sigma \rho=\frac{\sigma}{\chi^4\sqrt{1-\dot{\theta}^2}}.
 \label{eq2035}
\end{equation}
Neglecting $\dot{\theta}^2$ with respect to 1 we have
\begin{equation}
\rho_{\rm tach}\simeq  \frac{\sigma}{k^4}\frac{k^4}{\chi^4}.
 \label{eq3202}
\end{equation}
The value of the dimensionless parameter $\sigma/k^4$ may be estimated 
using the observational constraint on the amplitude of scalar perturbations.
Calculation of the power spectrum of scalar perturbations at the lowest order 
in $\epsilon_1$ and $\epsilon_2$ proceeds in the same way as in the standard tachyon
inflation \cite{steer} with the result
\begin{equation}
 \mathcal{P}_{{\rm S}} \simeq (0.44 + 2\alpha\epsilon_1+0.72 \epsilon_2)\frac{GH^2}{\pi\epsilon_1 }.
\label{eq3108} 
\end{equation}
Here  $\alpha$ is a parameter related to the expansion in $\epsilon_1$ of the speed of sound
$c_{\rm s}=1-2\alpha\epsilon_1 +O(\epsilon_1^2)$,
 which in our case yields $\alpha=1/12$.
For our purpose it is sufficient to use the  approximate expression $\mathcal{P}_{{\rm S}} \simeq GH^2/(\pi\epsilon_1)$
and compare with the power spectrum amplitude $A_s\simeq 2.2\times 10^{-9}$ measured by Planck 2015.
This implies a condition 
\begin{equation}
 \frac{H}{M_{\rm P}}=\frac{1}{\sqrt{24\pi}} \frac{\kappa^2 k^2\chi_{,\theta}^{1/2}}{\chi^2 \sqrt{\sigma}}
 \lesssim \sqrt{\pi A_s}\simeq 8.31\times 10^{-5} ,
\end{equation}
which must be satisfied close to and at the end of inflation (where $\epsilon_1 \lesssim 1$). 
Hence 
\begin{equation}
 \frac{\sigma}{k^4} \gtrsim \frac{10^{10}}{24 \pi \cdot 8.31^2}\frac{\kappa^4\chi_{,\theta}}{\chi^4}
 \label{eq01}
\end{equation}
and 
\begin{equation}
 \rho_{\rm tach} \gtrsim \frac{10^{10}}{24 \pi \cdot 8.31^2}\frac{\kappa^4\chi_{,\theta}}{\chi^8}k^4.
 \end{equation}
The tension of the D$_3$ brane is related to the string coupling constant $g_{\rm s}$ via
\cite{johnson}
\begin{equation}
 \sigma =\frac{1}{(2\pi)^3 \alpha^{\prime 2} g_{\rm s}},
 \label{eq02}
\end{equation}
where  $1/(2\pi\alpha^{\prime})$ is the string tension.
From this and (\ref{eq01}) we find a constraint 
\begin{equation}
 g_{\rm s}\lesssim \frac{3\cdot 8.31^2\cdot 10^{-10}}{\pi^2}
 \frac{\chi^4}{\kappa^4\chi_{,\theta}}\frac{M_{\rm s}^4}{k^4} ,
\end{equation}
where $M_{\rm s}=1/\sqrt{\alpha^{\prime}}$. 
Hence, with  $\kappa>1$ one can make the string coupling much less than unity even 
if $k\ll M_{\rm s}$
and we can choose $k$ and $M_{\rm s}$ such that the 
the natural scale hierarchy \cite{baumann}
\begin{equation}
 H <  M_{\rm s} < M_{\rm P}
 \label{eq03}
 \end{equation}
is satisfied.
%With this in mind, for any given function $\chi(\theta)$ we can estimate a lower bound on   
%our free parameter $\kappa$ the value of which
% is related to $\chi(\theta_{\rm f})$ via (\ref{eq017}).
%A lower bound is obtained by combining equations (\ref{eq03}) and (\ref{eq017})  with 
%the requirement 
%$H\lesssim k$, i.e., $h\lesssim 1$.

To estimate the proportion of radiation at the end of inflation we will use the approximate expression (\ref{eq2015})
 in the  frequency interval $ F <\omega <\infty$ and neglect the contribution 
 in the interval $0< \omega < F$.
 In this way we obtain an  estimate of the integral
(\ref{eq2031}) 
\begin{equation}
 \rho_{\rm rad}(t) \simeq
 \frac{F^4}{16\pi^2}\int_F^\Lambda \frac{d\omega}{\omega}
 =\frac{F^4}{16\pi^2} \ln \frac{\Lambda}{F} ,
  \label{eq2032}
\end{equation}
where $\Lambda$ is a physical momentum cutoff of the order of $k$.
This expression should be compared with the tachyon energy density
(\ref{eq3202}).
To estimate $F$ we use 
Eq. (\ref{eq2016}) which can be written as   
\begin{equation}
F^2=k^2\left|
 \frac32  h \frac{\dot{f}}{f}+\frac12 \frac{\ddot{f}}{f}-\frac14\left(\frac{\dot{f}}{f}\right)^2
 \right| .
\label{eq2020}
\end{equation}
From (\ref{eq2036}) we find 
\begin{equation}
 \dot{f}=\frac{2\dot{\theta}^3\ddot{\theta}}{f},
 \quad
 \ddot{f}=\frac{6\dot{\theta}^2\ddot{\theta}^2+2\dot{\theta}^3\dddot{\theta}}{f}
 -\frac{4\dot{\theta}^6\ddot{\theta}^2}{f^3}
 \label{eq2037}
\end{equation}
and calculate $h$ using Eq.\ (\ref{eq1007}).

\subsection{Reheating in the standard tachyon inflation}
\label{standard}
For the sake of comparison we first make an estimate of the reheating in the tachyon
inflation of the standard cosmology. In this case for 
$\chi=\theta^n $ we find
\begin{equation}
h=\frac{\kappa}{\sqrt{3}}\frac{1}{\chi^2} ,
\end{equation}
\begin{equation}
\dot{\theta}=\frac{4}{\sqrt{3}\kappa}\chi_{,\theta}\chi ,
\label{eq2101}
\end{equation}
\begin{equation}
\ddot{\theta}=\frac{16}{3\kappa^2}\big(\chi_{,\theta\theta}\chi_{,\theta}\chi^2+\chi_{,\theta}^3\chi\big) ,
\label{eq2102}
\end{equation}
\begin{equation}
\dddot{\theta}=\frac{64}{3\sqrt{3}\kappa^3}\big(\chi_{,\theta\theta\theta}
\chi_{,\theta}^2\chi^3+\chi_{,\theta\theta}^2\chi_{,\theta}\chi^2+
5\chi_{,\theta\theta}\chi_{,\theta}^3\chi^2+\chi_{,\theta}^5\chi\big) .
\label{eq2103}
\end{equation}
Furthermore we also have 
\begin{equation}
\dot{h}=-\frac83\frac{\chi_{,\theta}^2}{\chi^2} ,
\end{equation}  
yielding
\begin{equation}
\epsilon_1=\frac{8\chi_{,\theta}^2\chi^2}{\kappa^2}.
\end{equation}
From this and the condition $\epsilon_1\simeq 1$ at the end of inflation we obtain  
\begin{equation}
\chi_{\rm f}^2=\bigg(\frac{\kappa^2}{8n^2}\bigg)^{n/(2n-1)} ,
\end{equation}
\begin{equation}
h_{\rm f}=\frac{\kappa}{\sqrt{3}}\bigg(\frac{\kappa^2}{8n^2}\bigg)^{-n/(2n-1)} ,
\end{equation}
where by $h_{\rm f}$ we denote 
and $h(\theta_{\rm f})$.
Then, using (\ref{eq2101})-(\ref{eq2103}) we find
\begin{eqnarray}
\dot{\theta_{\rm f}}=\sqrt{\frac{2}{3}},\qquad\ddot{\theta_{\rm f}}=\frac{2n-1}{\sqrt{6}n}h_{\rm f},\qquad
\dddot{\theta_{\rm f}}=\frac14\sqrt{\frac{2}{3}}\frac{(4n-3)(2n-1)}{n^2}h_{\rm f}^2.
\end{eqnarray}
From this and Eqs.\ (\ref{eq2020}) and (\ref{eq2037}) we obtain 
\begin{equation}
F=\frac{|2n-1|^{1/2}|46n-17|^{1/2}}{2\cdot 3^{3/2}n}H_{\rm f},
\end{equation}
where $H_{\rm f}\equiv h_{\rm f}k$ is the physical Hubble rate at the end of inflation.
It is worth noting that the cosmological particle creation vanishes exactly for $n=1/2$.
This power precisely equals 
the critical power at which the tachyon cosmology changes
from dust to quasi-de Sitter \cite{abramo,bilic4}. 

To estimate the tachyon energy density we can use the expression (\ref{eq2035}) with $\sigma$ estimated by
the inequality (\ref{eq01}) in which $\chi_{,\theta}$ is set to 1
yielding
\begin{equation}
\rho_{\rm tach}\gtrsim\frac{9\cdot10^{10}}{24\pi(8.31)^2}H_{\rm f}^4.
\end{equation}
Finally we obtain the estimate if the ratio
\begin{equation}
\frac{\rho_{\rm rad}}{\rho_{\rm tach}}\lesssim\frac{(2n-1)^2(46n-17)^2(8.31)^2}{2^53^7n^4\pi}\cdot10^{-10} .
\end{equation}
Hence,  the cosmologically created radiation in the standard tachyon cosmology is negligible.

\subsection{Reheating in the BWC tachyon inflation}
\label{bwc}
Next we proceed by estimating the radiation density in the BWC tachyon inflation. 
At the end of inflation we can neglect the quartic term 
$\kappa^2/\chi_{\rm f}^4$  with respect to $\chi_{,\theta}$. 
Then, using Eqs. (\ref{eq0021}) and (\ref{eq0018}) and the condition
(\ref{eq017}) at the end of inflation we find
\begin{equation}
h_{\rm f}=\frac{\kappa}{\sqrt3}\frac{\chi_{{\rm f},\theta}^{1/2}}{\chi_{\rm f}^2},
\quad\quad
 \dot{\theta}_{\rm f}\simeq \frac{4}{\sqrt{3}} \frac{\chi_{\rm f}^{}\chi_{{\rm f},\theta}^{1/2}}{ \kappa},
\quad
\ddot{\theta}_{\rm f}\simeq\frac{16\chi_{\rm f}^{}\chi_{{\rm f},\theta}^2}{\kappa^2} 
\left(1-\frac{1}{12}
\frac{\kappa^2}{\chi_{\rm f}^2\chi_{{\rm f},\theta}^{}}
\right),
 \label{eq3001}
\end{equation}
\begin{equation}
\dddot{\theta_{\rm f}}\simeq 
\frac{32}{3\sqrt{3}}\frac{\chi_{\rm f}^{}\chi_{{\rm f},\theta}^{7/2}}{\kappa^3}
\left(26
-\frac {3\kappa^2}{\chi_{\rm f}^2\chi_{{\rm f},\theta}^{}}
+\frac{\chi_{\rm f}^2\chi_{{\rm f},\theta\theta\theta}^{}}{\chi_{{\rm f},\theta}^3}
\right).
 \label{eq3013}
\end{equation}

Consider first the exponential potential $V=\sigma e^{-\theta}$. , i.e.,  $\chi(\theta)=e^{\theta/4}$. 
Then, from (\ref{eq017}) we find 
\begin{equation}
 \chi_{\rm f}^3=\frac23 \kappa^2.
 \label{eq5001}
\end{equation}
Using this and (\ref{eq3107}) in the limit $n\rightarrow\infty$
it follows
\begin{equation}
h=\frac{1}{\sqrt8}
\end{equation}
and from (\ref{eq5006})
we obtain
\begin{equation}
\rho_{\rm rad}\simeq \frac{10^4 k^4}{ 3^{14}\pi^2} 
\ln \frac{\Lambda}{F}=\simeq 0.01356 H_{\rm f}^4  .
 \label{eq5003}
\end{equation}
This has to be compared with the tachyon energy density
\begin{equation}
\rho_{\rm tach}\simeq \frac{\sigma}{\chi_{\rm f}^4} \gtrsim \frac{3\cdot 3^{1/3}
\cdot 10^{10}}{2^7\cdot 2^{1/3}\cdot 8.31^2\pi}\frac{k^4}{\kappa^{2/3}},
 \label{eq5002}
\end{equation}
yielding an estimate of the ratio
\begin{equation}
\frac{\rho_{\rm rad}}{\rho_{\rm tach}}\lesssim\frac{2^{12}\cdot 5^4
\cdot 8.31^2\cdot 10^{-10} }{3^{15}\cdot 3^{1/3} \pi} 
\left(\frac{\kappa}{2}\right)^{2/3}
=2.719 \cdot 10^{-10} \left(\frac{\kappa}{2}\right)^{2/3} .
 \label{eq3103}
\end{equation}
Since the parameter $\kappa$ can can be arbitrary large 
this ratio can, in principle, be made close to unity. However this would require unnaturally large
value of $\kappa$.

Next, consider  the general power law potential, i.e.,  $\chi(\theta)=\theta^n$.  
In this case we find 
\begin{equation}
h_{\rm f}=\left(\frac{n}{6 (3n+1)}\right)^{1/2}\frac{\kappa^2}{\chi_f^3},
\label{eq3107}
\end{equation}
\begin{equation}
  \dot{\theta}_{\rm f}= \left(\frac{8n}{3(3n+1)}\right)^{1/2},
\end{equation}
\begin{equation}
  \ddot{\theta}_{\rm f}=\frac{2n(3n-1)}{3(3n+1)^2}\frac{\kappa^2}{\chi_f^3},
\end{equation}
\begin{equation}
 \dddot{\theta_{\rm f}}=\frac{4n^{3/2}(3n-1)(3n-2)}{3\sqrt6(3n+1)^{7/2}} 
  \left(\frac{\kappa^2}{\chi_f^3}
  \right)^2 .
  \label{eq3109}
\end{equation} 
 Using this and equations (\ref{eq2020}) and (\ref{eq2037})
 we find 
 \begin{equation}
F= \frac{2^3n|3n-1|^{1/2} |150n^3+85n^2-3|^{1/2}}{3^{3/2}(3n+1)^3} H_{\rm f}
%\ln \frac{\Lambda}{F}
 \label{eq5004}
\end{equation} 
and from (\ref{eq2032}) we find
 \begin{equation}
\rho_{\rm rad}\simeq \frac{2^8n^4(3n-1)^2(150n^3+85n^2-3)^2}{ 3^{6}\pi^2(3n+1)^{12}} H_{\rm f}^4
\ln \frac{\Lambda}{H_{\rm f}}.
 \label{eq5006}
\end{equation} 
Note that the limit $n\rightarrow\infty$
 the righthand sides of (\ref{eq3107})-(\ref{eq5006}) approach finite nonzero values
  corresponding to the exponential potential $V=\sigma e^{-\theta}$. 
It is worth mentioning that the radiation density described by Eq. (\ref{eq5006}) is, up to the multiplicative constant,
equal to that obtained in the conventional calculation of  particle creation 
  \cite{ford}.
Note that the cosmological particle creation vanishes exactly for $n=1/3$. 
Again, this power 
precisely equals
the critical power at which the tachyon cosmology changes
from dust to quasi-de Sitter \cite{bilic4}.

%plot:
%black:   (2x-2)/(3x-1)*log(2)/log(10)-10
%red:     (2x-2)/(3x-1)*log(5)/log(10)-10
%blue:     (2x-2)/(3x-1)*log(15)/log(10)-10

\begin{figure*}[ht]
\begin{center}
\includegraphics[width=0.7\textwidth,trim= 0 0 0 0]{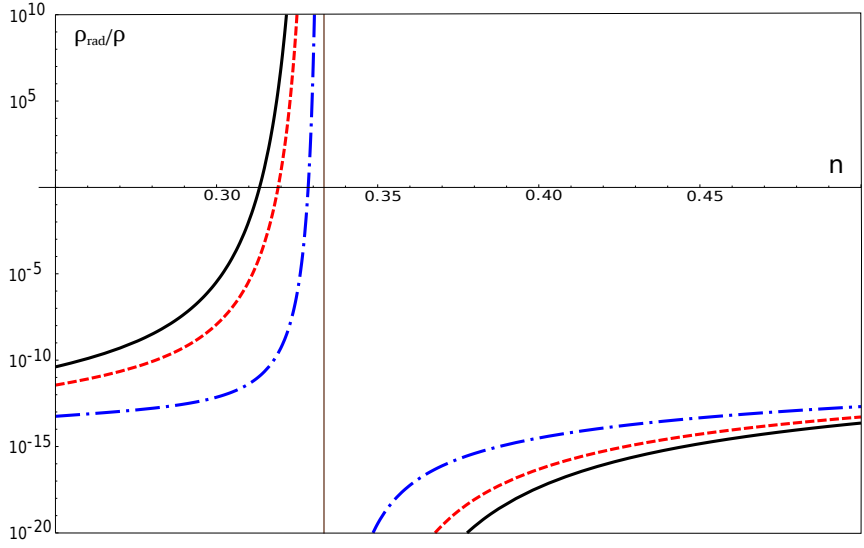}
\caption{$\rho_{\rm rad}/\rho_{\rm tach}$ as a function of $n$ for the power law 
tachyon potential $V=\theta^{-4n}$ for $\kappa=$ 3 (full black line), 6 (dashed red),
and 9 (dash-dotted blue). The vertical line indicates the critical power $n=1/3$.
}
\label{fig1}
\end{center}
\end{figure*}

From (\ref{eq017}) we find
\begin{equation}
 \chi_{\rm f}^{3-1/n}=\frac{\kappa^2}{2(3n+1)},
 \label{eq5007}
\end{equation}
yielding
\begin{equation}
h_{\rm f}=\sqrt{\frac{2n(3n+1)}{3}}\left(\frac{2(3n+1)}{\kappa^2} \right)^{1/(3n-1)} .
\label{eq3106}
\end{equation}
Using this and (\ref{eq5006}) we obtain
\begin{equation}
\rho_{\rm rad}=\frac{C_n (3n-1)^2 n^6 k^4}{24\pi (8.31)^2}(3n+1)^{\frac{7n+1}{3n-1}}2^{\frac{5n+3}{3n-1}}
\kappa^{-\frac{8}{3n-1}},
\end{equation}
where the coefficient
\begin{equation}
C_n =\frac{2^{12}\cdot 8.31^2}{3^7\pi (3n+1)^{13} }\left(\frac{3n+1}{2}\right)^{2n/(3n-1)}
\left(150n^3+85n^2-3\right)^2
%C_n \cdot 10^{-10} \kappa^{2/3-2/(9n-3)} 
 \label{eq3105}
\end{equation}
%
%C_n= 2^12* 8.31^2/(3^7 *pi* (3*x+1)^13)*((3*x+1)/2)^(2*x/(3*x-1))*(150*x^3+85*x^2-3)^2
%
is a smooth function of  $n$ for $n\geq 0$ assuming the maximal value of 1.018447 at $n=0.316770$.

For the tachyon energy density we obtain the lower bound
\begin{equation}
\rho_{\rm tach}\gtrsim\frac{10^{10}nk^4}{24\pi(8.31)^2}(3n+1)^{\frac{7n+1}{3n-1}}2^{\frac{7n+1}{3n-1}}
\kappa^{-\frac{2n+6}{3n-1}},
\end{equation}
yielding an estimate of the ratio
\begin{equation}
\frac{\rho_{\rm rad}}{\rho_{\rm tach}}\lesssim
C_n n^5(3n-1)^2\cdot 10^{-10} \left(\frac{\kappa}{2}\right)^{(2n-2)/(3n-1)} .
 \label{eq3104}
\end{equation}
Note that for $\kappa>2$, the reheating is enhanced for $n>1$ and $ n < 1/3 $. In the limit
$n\rightarrow \infty$  the power of $\kappa$ approaches 2/3 yielding the enhancement as in the exponential case.
In the limit
 $n\rightarrow 1/3$ from below the righthand side of (\ref{eq3104})  diverges, hence 
the ratio $\rho_{\rm rad}/\rho_{\rm tach}$can be arbitrary large for $n$ sufficiently close to $1/3$.
Thus, in order to obtain a significant enhancement we need $\kappa>2$ and $n$ below and close to 1/3.
The inequality $\kappa>2$ implies
 \begin{equation}
 h_{\rm f} \gtrless \sqrt{\frac{2n(3n+1)}{3}}\left(\frac{(3n+1)}{2} \right)^{1/(3n-1)}  \quad {\rm for}\quad
 n\lessgtr 1/3 .
 \end{equation}
% 
% kappa > f=(2*(3x+1))^(1/2) (2/3*x*(3x+1))^((3x-1)/4)
%where x=n
% 
%f'= (2/(x+2)+x/(x+1)+x/(x+2)+ln(x+1)+ ln(x+2)+ln(2/9))*(x+2)^2*((2x+2)*(x+2)/9)^x
%where x=3n-1
%h=(2x*(3x+1)/3)^(1/2)*(2(3x+1)/y^2)^(1/(3x-1))
% where x=n, y=kappa
%
%h< ff=(2x*(3x+1)/3)^(1/2)*((3x+1)/2)^(1/(3x-1))
The right hand side of this inequality
is a monotonously increasing function of $n$ 
taking the values of 0.921321, 1,  1.099148,  at
$n$ equal to 1/4, 0.286106, and 1/3, respectively.
Roughly, this means that a significant enhancement
requires $h_{\rm f}>1$ to wit $H_{\rm f}>k$. 
In figure \ref{fig1} we plot the ratio $\rho_{\rm rad}/\rho_{\rm tach}$ as a function of $n$ for various 
values of $\kappa>2$.

  % log(rho_rad/rho=
%-11.872857872+2x/(3x-1)log((3x+1)/2)+5log(x)-13log((3x+1)/2)+2log(150x^3+85x^2-3)+log((3x-1)^2)+(2x-2)/(3x-1)log(k/2)
 %kappa=3  black #000000
 %kappa=6 red #FA0004
 %kappa=9 blue #002FFF
%\newpage
%\appendix
%$$
%-11.872857872+2x/(3x-1)log((3x+1)/2)+5log(x)-13 log((3x+1)/2)
%$$
%$$
%+2log(150x^3+85x^2-3)+log((3x-1)^2)+ (2x-2)/(3x-1)log(k/2)
%$$
\section{Conclusions}
\label{conclude}
We have investigated the reheating in a braneworld inflationary scenario based on
 coupling of the tachyon with the abelian gauge field and the 
 cosmological creation of massless particles.
 Assuming the  tachyon potential of the inverse power $V\propto \theta^{-4n}$ 
 we have shown that the cosmological creation of massless particles vanishes for critical power
 $n=1/2$ in the standard cosmology and $n=1/3$ in BWC.
Next, we have shown that the reheating due to cosmological particle creation 
is insignificant in the standard cosmology whereas
in BWC
the reheating depends strongly on the
 power and can be significantly enhanced for powers $n$ approaching a critical point 
$n=1/3$ from below.

Unfortunately this scenario alone cannot solve the reheating problem
of the tachyon inflation.
It has been  shown \cite{abramo,bilic4} that
the energy density of the tachyon with an inverse power potential yields  asymptotically either 
dust or quasi de sitter universe, with  the cosmological scale dependence as 
$\rho_{\rm tach}\propto a^{-3}$ or $1/\log a$, respectively.
Since the radiation density behaves as $\rho_{\rm rad} \propto a^{-4}$,
sooner or later $\rho_{\rm tach}$ will inevitably dominate the radiation.

It would be of considerable interest to investigate the effects of
cosmological creation in the warm inflation models \cite{berera}.
In warm inflation, radiation due to dissipative effects
is produced in parallel with the inflationary
expansion and  inflation ends when the universe heats up to become radiation dominated.
This scenario has been successfully applied to tachyon inflation models
\cite{herrera,motaharfar} and, in principle, should also work for
tachyon inflation in BWC presented here.

\section*{Acknowledgments}

This work has been supported by the Croatian Science Foundation under the project 
IP-2014-09-9582 and partially supported by ICTP - SEENET-MTP project NT-03 
Cosmology - Classical and Quantum Challenges.
N.\ Bili\'c and S.\ Domazet 
were partially supported by the H2020 CSA Twinning project No.\ 692194, ``RBI-T-WINNING''.
 G. Djordjevic acknowledges support
by the Serbian Ministry for Education, Science and Technological Development under the
project No. 176021 and by CERN-TH Department.

\end{document}